\documentclass[10pt]{article}
\usepackage{graphicx}
\usepackage{amsmath}

\def\Title#1{\begin{center} {\Large #1 } \end{center}}
\def\Author#1{\begin{center}{ \sc #1} \end{center}}
\def\Address#1{\begin{center}{ \it #1} \end{center}}

\newcommand\pubblock{\rightline{\begin{tabular}{l} Proceedings of the Second Annual LHCP\\ \pubnumber\\
         \pubdate  \end{tabular}}}

\newenvironment{Abstract}{\begin{quotation} \begin{center} 
             \large ABSTRACT \end{center}\bigskip 
      \begin{center}\begin{large}}{\end{large}\end{center} \end{quotation}}

\newenvironment{Presented}{\begin{quotation} \begin{center} 
             PRESENTED AT\end{center}\bigskip 
      \begin{center}\begin{large}}{\end{large}\end{center} \end{quotation}}

\def\beq{\begin{equation}}
\def\eeq#1{\label{#1}\end{equation}}
\def\eeqn{\end{equation}}

\def\beqa{\begin{eqnarray}}
\def\eeqa#1{\label{#1}\end{eqnarray}}
\def\eeqan{\end{eqnarray}}

\let\bar=\overbar

\def\Dslash{\not{\hbox{\kern-4pt $D$}}}
\def\dslash{\not{\hbox{\kern-2pt $\del$}}}

\def\msb{{\bar{\ssstyle M \kern -1pt S}}}

\usepackage{color}

\newcommand\pubnumber{ CMS CR-2014/168 }

\newcommand\pubdate{\today}

\def\affiliation{
On behalf of the CMS Experiment, \\
Institute for Particle Physics, \\
ETH Zurich, Switzerland }

\begin{document}

\large
\begin{titlepage}
\pubblock

\vfill
\Title{  Measurement of VZ production cross sections in Z$\rightarrow \rm{b\bar{b}}$ decay channels in pp collisions at 8 TeV  }
\vfill

\Author{ Philipp Eller  }
\Address{\affiliation}
\vfill
\begin{Abstract}

We present a measurement of the WZ and ZZ production cross sections in proton-proton collisions at 8 TeV in final states where one Z boson decays to b-tagged jets, while the other gauge boson, either W or Z, is detected through its leptonic decay. The results are based on data corresponding to an integrated luminosity of 18.9 fb$^{-1}$ collected with the CMS detector at the Large Hadron Collider.

\end{Abstract}
\vfill

\begin{Presented}
The Second Annual Conference\\
 on Large Hadron Collider Physics \\
Columbia University, New York, U.S.A \\ 
June 2-7, 2014
\end{Presented}
\vfill
\end{titlepage}
\def\thefootnote{\fnsymbol{footnote}}
\setcounter{footnote}{0}
%

\normalsize 


\section{Introduction}

Measurements are reported of the WZ and ZZ production cross sections in proton-proton collisions at √s = 8 TeV in final states where one Z boson decays to b-tagged jets. The other gauge boson, either W or Z, is detected through its leptonic decay. The results are based on data corresponding to an integrated luminosity of 18.9 fb$^{-1}$ collected with the CMS detector \cite{Chatrchyan:2008aa} at the Large Hadron Collider. This measurement has recently been published \cite{SMP-13-011}.

\section{Analysis}

The measurement presented here is a spin-off of the CMS analysis of the asociated Higgs production, where the Higgs decays into b-quarks \cite{Chatrchyan:2013zna} and employs the same techniques.
Events are categorised depending on the leptonic decay of the W or Z boson (either 
$\rm{W \rightarrow e\nu,\ \mu\nu}$ or 
$\rm{Z \rightarrow e^+e^−,\ \mu^+\mu^−,\ or\ \nu\bar{\nu}}$) 
and its transverse momentum.
One key feature is the regression that is used to better estimate the true b-
energy, improving the di-jet mass resolution up to 15\%. 
This technique employs a boosted decision tree (BDT) using jet properties as inputs  to assign a correction to the jet energy. The improvement  is illustrated in the figure.

The major background processes tt-bar and drell-yan + light and heavy flavoured jets are normalized to data in mass sidebands. The other backgrounds from single top production and associated Higgs production are normalized to theory calculations. Most important systematic uncertainties are treated either as scale uncertanties, i.e. uminosity, lepton efficiencies and triggerand background normalizations, or taken into account as shape variations for jet energy scale and resolution, b-tagging , MC statistics and modeling. Their effect on the signal stregth $\mu$ are summarized in  Table \ref{tab:table1}.

\begin{figure}[htb]
\centering
\includegraphics[height=2in]{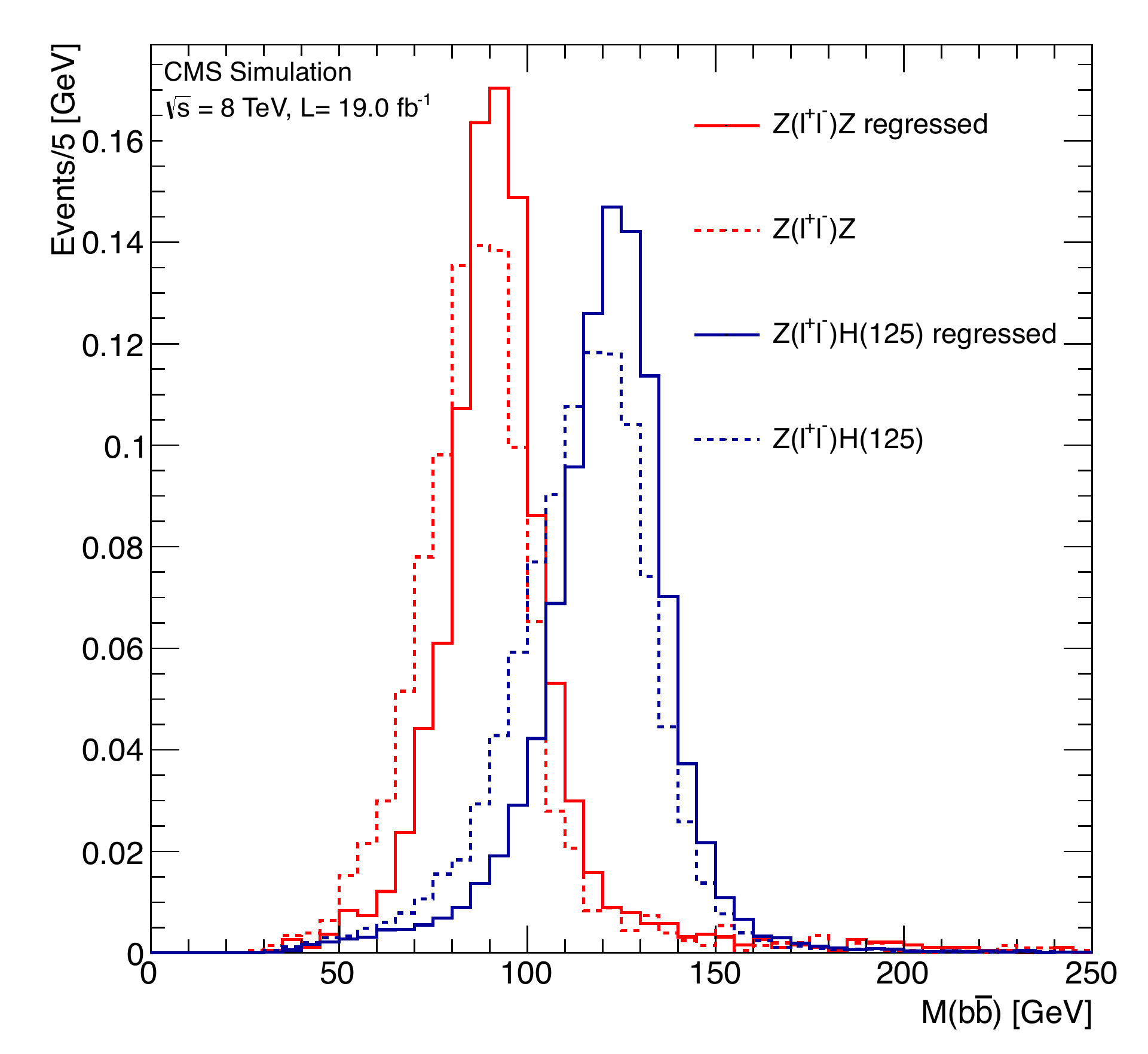}
\caption{ Regressed vs. non-regressed mass of the di-jet system from the Z decay (red) and the H decay (blue) for the analysis in \cite{Chatrchyan:2013zna} }
\label{fig:figure1}
\end{figure}

\begin{figure}[htb]
\centering
\includegraphics[height=2in]{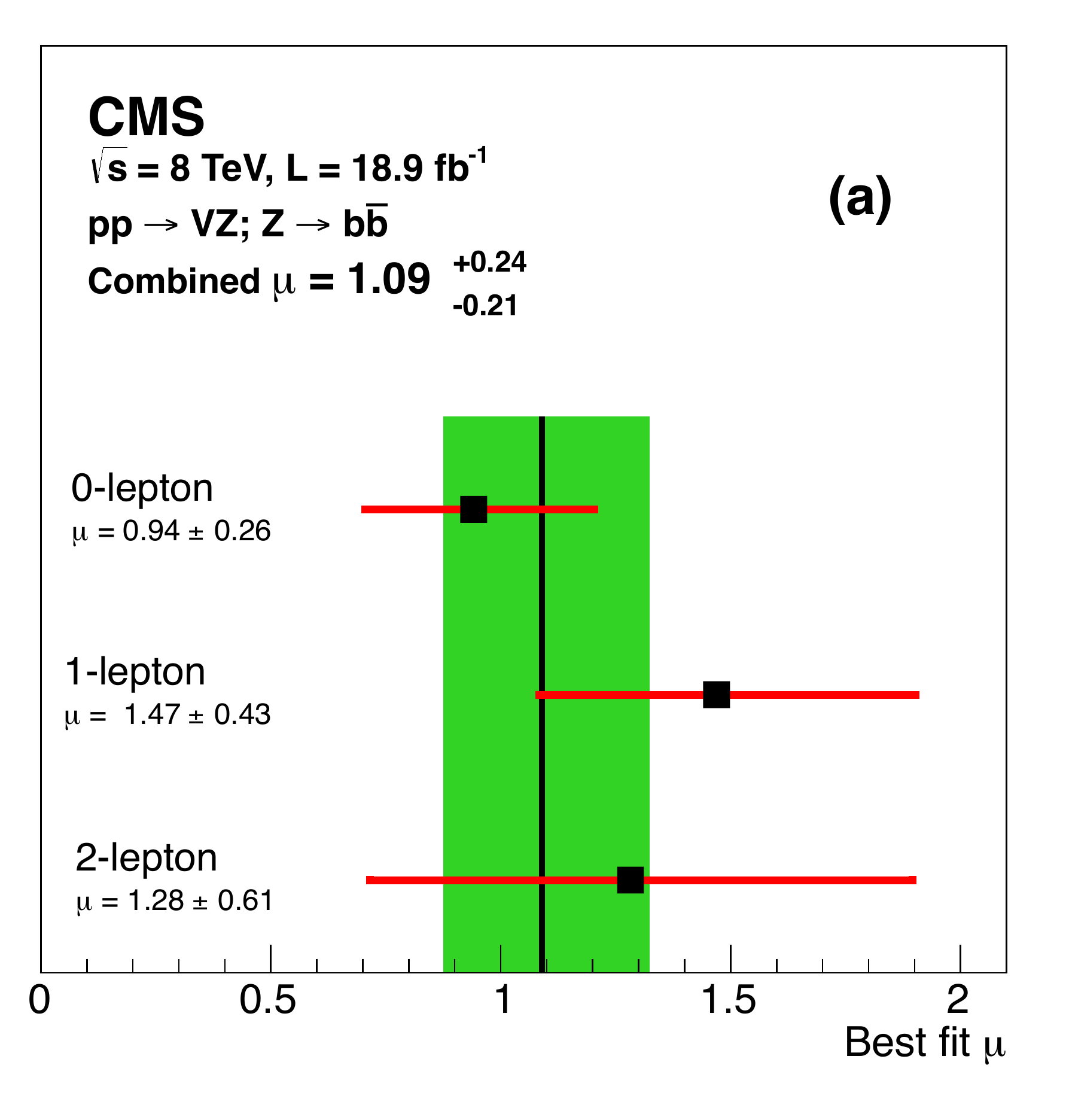}
\includegraphics[height=2in]{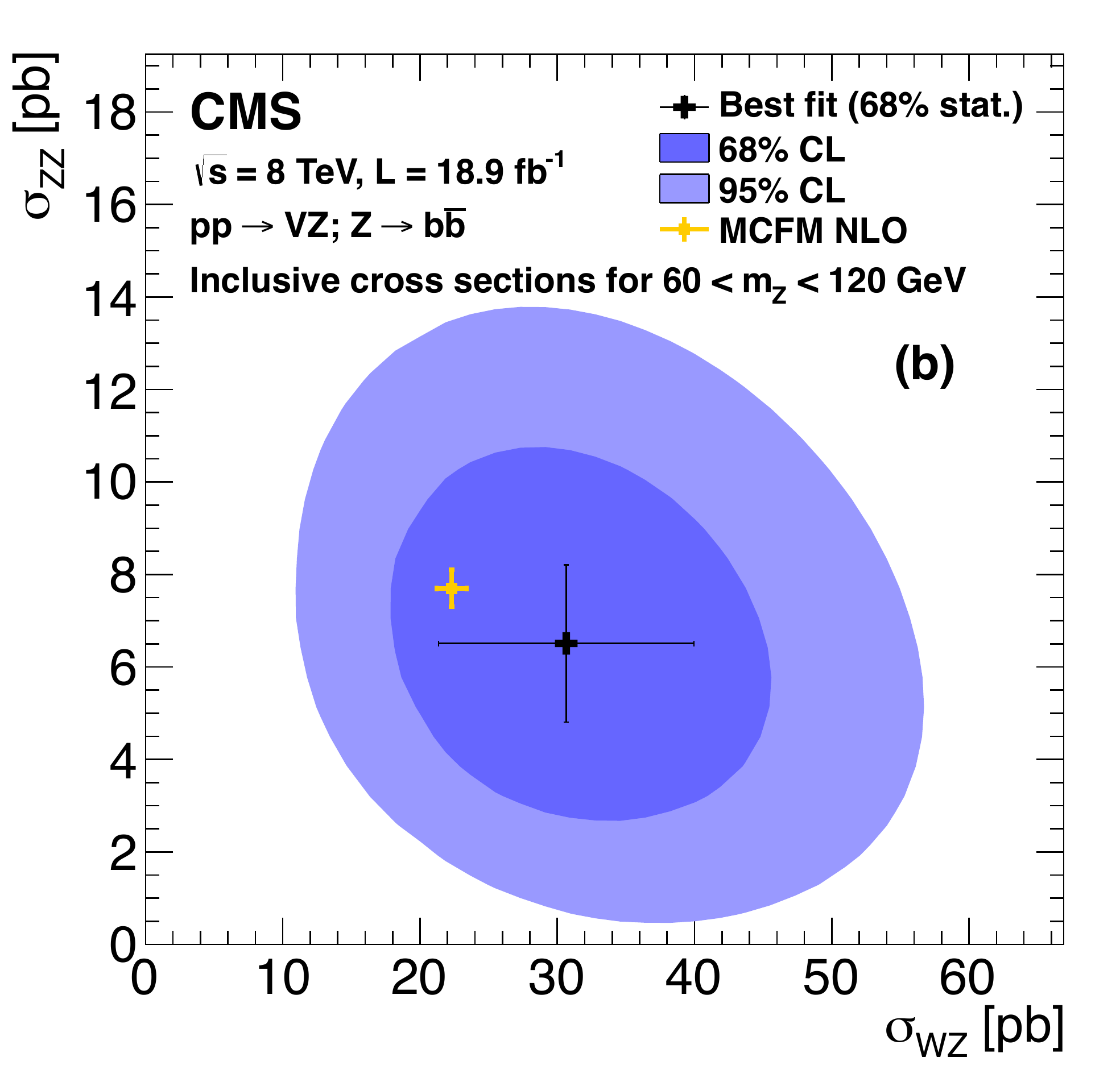}
\caption{ (a) Best-fit values of the ratios of the VZ production cross sections, relative to SM predictions for individual channels, and for all channels combined (green band). (b) Contours of 68\% and 95\% confidence level for WZ and ZZ production cross sections.}
\label{fig:figure2}
\end{figure}

\begin{figure}[htb]
\centering
\includegraphics[height=2in]{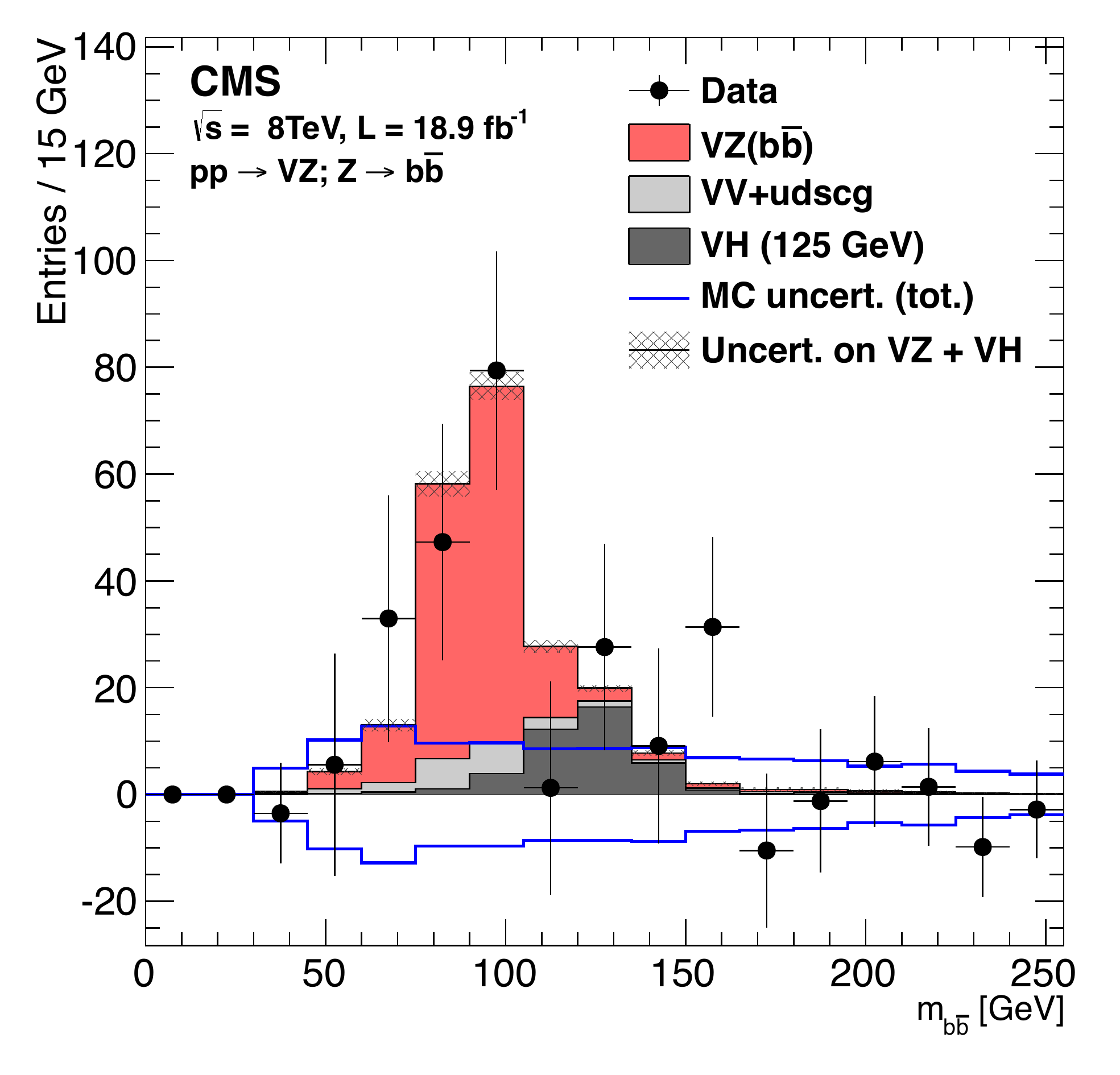}
\caption{ The combined bb invariant mass distribution for all channels, compared to MC simulation of SM contributions with all backgrounds to VZ production subtracted, except for the VH contribution.}
\label{fig:figure3}
\end{figure}

\begin{figure}[htb]
\centering
\includegraphics[height=2in]{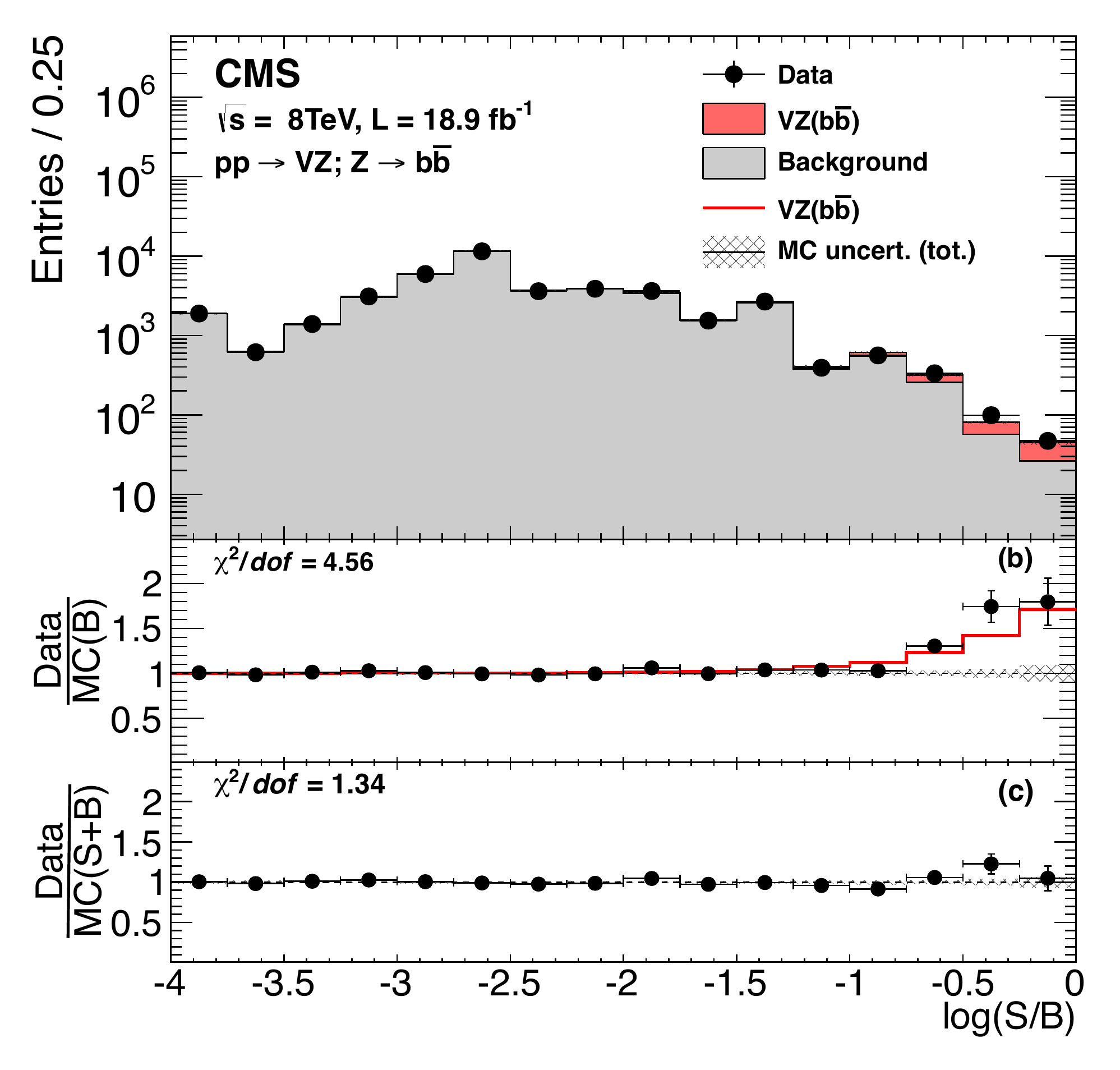}
\caption{ Combined distribution of the BDT output for all channels sorted in bins 
of signal to background ratios in data and in Monte Carlo (MC) simulations and the ratios of data to background and signal + background.}
\label{fig:figure4}
\end{figure}

\begin{figure}[htb]
\centering
\includegraphics[height=2in]{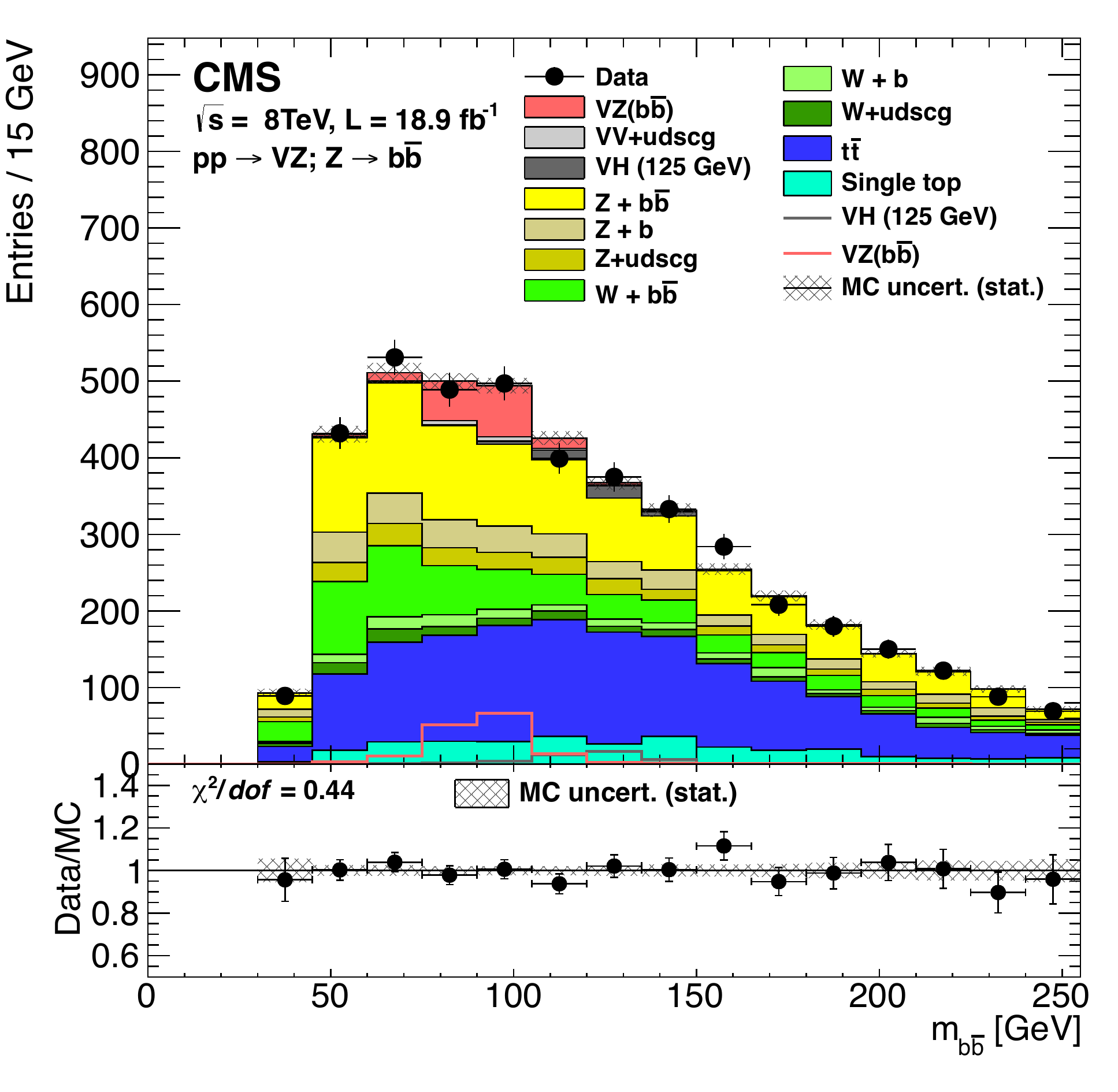}
\caption{ The combined di-jet invariant mass distribution for all channels, compared to MC simulation of SM contributions.}
\label{fig:figure5}
\end{figure}



\begin{table}[t]
\begin{center}
\begin{tabular}{r|c}  
& Uncertainty on $\mu$  \\ \hline
Luminosity & 3\% \\
lepton efficiencies and trigger & 2\% \\
background normalizations &2-13\% \\
jet energy scale &7\% \\
jet energy resolution &6\% \\
b-tagging &7\% \\
MC statistics &5\% \\
MC modeling both &5\% \\ \hline
\end{tabular}
\caption{ most important sources of systematic uncertaintiy on the signal stregth  ($\mu$)}
\label{tab:table1}
\end{center}
\end{table}

\section{Results}

Signal extraction is performed by two different methods: a Multivariate technique using a BDT to best separate signal from background and a more direct analysis on the invariant mass spectrum of the di-jet system. Following the numbers based on the BDT are given.
The measured cross sections for both processes are
\begin{equation*}
\rm{\sigma(pp \rightarrow WZ) = 30.7 \pm 9.3\ (stat.) \pm 7.1\ (syst.) \pm 4.1\ (th.) \pm 1.0\ (lum.)\ pb }
\end{equation*}
\begin{equation*}
\rm{\sigma(pp\rightarrow ZZ) = 6.5 \pm 1.7\ (stat.) \pm 1.0\ (syst.) \pm 0.9\ (th.) \pm 0.2\ (lum.)\ pb}
\end{equation*}
and are consistent with SM NLO QCD calculations.
Signal strength values for VZ production are consistent among all three event categories (0,1 and 2 lepton channels) and consistent with SM predictions.


\begin{thebibliography}{99}


\bibitem{Chatrchyan:2008aa}
  S.~Chatrchyan {\it et al.}  [CMS Collaboration],
  JINST {\bf 3} (2008) S08004.

\bibitem{SMP-13-011}
 S.~Chatrchyan {\it et al.}  [CMS Collaboration],
Eur. Phys. J. C (2014) 74:2973.

\bibitem{Chatrchyan:2013zna}
  S.~Chatrchyan {\it et al.}  [CMS Collaboration],
  Phys.\ Rev.\ D {\bf 89} (2014) 012003
  [arXiv:1310.3687 [hep-ex]].


\end{thebibliography}
\end{document}